\documentclass{PoS}

\title{Recent developments in jet quenching theory}

\ShortTitle{Recent developments in jet quenching theory}

\author{\speaker{Liliana Apolin\'{a}rio}\\
        Departamento de F\'{i}sica de Part\'{i}culas and IGFAE, Universidade de Santiago de Compostela, \\15706 Santiago de Compostela, Galicia-Spain\\
        CENTRA, Instituto Superior T\'{e}cnico, Universidade T\'{e}cnica de Lisboa, \\Av. Rovisco Pais, P-1049-001 Lisboa, Portugal \\
        E-mail: \email{lilianamarisa.cunha@usc.es}}

\abstract{Motivated by the new results obtained in heavy-ion collision experiments at the LHC, several extensions of the standard calculations of energy loss have been made recently. In this manuscript, I provide a short overview of some of the recent developments in jet quenching theory. First, I discuss some improvements computed by different groups to implement energy-momentum conservation in a rigorous way, relaxing some of the assumptions that were made in the standard jet quenching calculations. Second, quantum interference effects between different parton emitters when propagating through an extended coloured medium will be considered, with a quark-antiquark antenna as a model setup. Finally, other recent extensions as the modification of the colour flow inside a jet in a QCD medium with respect to vacuum, and the use of SCET to compute transverse momentum broadening and medium-induced gluon radiation, will be very briefly presented.}

\FullConference{The European Physical Society Conference on High Energy Physics -EPS-HEP2013\\
		18-24 July 2013\\
		Stockholm, Sweden}

\begin{document}

\section{Introduction}
\label{sec1}

\par The study of jet quenching phenomena is at the forefront of the current theoretical and experimental efforts in heavy-ion collisions. Its quantitative understanding is essential for an accurate characterization of the hot and dense medium that is produced in the collisions, the quark-gluon plasma (QGP). To accomplish this goal, one of the main objectives is the determination of the main mechanisms of in-medium energy loss and the consequent evaluation of the medium-induced modifications to the evolution of the parton shower. 
\par The standard jet quenching description before nucleus-nucleus collisions at the LHC was made in terms of radiative energy loss only. Several formalisms were built (see \cite{models1,models2,models3} and references therein), all of them based on a perturbative description that assumes a factorization of multiple gluon emissions as a single collinear gluon emission process. Additionally, most of them work in the limit of soft gluon emissions, $x \rightarrow 0$, with $x$ being the fraction of energy carried by the gluon. All these assumptions raise several problems and limitations, that are discussed at length in \cite{models2}. The fact that the region of validity is different for each model and the phenomenological extensions into the large $x$ and large angle domain that are necessary for a complete description of the process, lead to large discrepancies between the formalisms. Furthermore, the shower evolution is based on multi-gluon radiation and therefore the problem of colour coherence between different emitters should also be addressed. In most of the models, the hadronization is assumed to be unmodified by the presence of a medium. However, colour coherence or decoherence effects between the different emitters can induce modifications independently of where the hadronization takes place. 
\par In this manuscript, I address some of the efforts that were made to improve the limitations mentioned before. For that, in section \ref{sec2}, several works within the path-integral formalism that calculate the single emission process in a more general kinematic range will be presented. This kind of calculations is also being explored within the Soft-Collinear Effective Theory (SCET), that will be summarily addressed in section \ref{sec3}. In section \ref{sec4}, it will be shown the phenomenological consequences of the latest developments of the QCD antenna setup considering the propagation of colour correlated sources inside a coloured medium. A recent study of medium-modified colour connections will be presented in section \ref{sec5}. For the latest developments within the lattice QCD description, see the proceedings of this conference \cite{panero1} and the work of \cite{lattice1,lattice2}. A final summary will be presented in section \ref{sec6}.

\section{Medium-induced radiation within the path-integral formalism} 
\label{sec2}

Among the several energy loss mechanisms, the radiative channel is the dominant process of energy loss of a highly enough energetic parton that undergoes multiple soft scattering with the medium constituents. The elementary process under consideration is the propagation of a highly energetic quark through a medium with longitudinal boundaries at $[x_{0+}, L_+]$, that is described by a Wilson line:
\begin{equation}
\label{eq:Wline}
	W (x_{0+}, L_+; \mathbf{x}) = \mathcal{P} \exp \left\{ ig \int_{x_{0+}}^{L_+} dx_+ A_- (x_+, \mathbf{x}) \right\} ,
\end{equation}
where $A_- (x_+, \mathbf{x})$ are the colour fields of the scattering center located at $x_+$ with transverse coordinate $\mathbf{x}$ that will induce a colour rotation of the propagating projectile. These fields are integrated in a path-ordered way identified by $\mathcal{P}$. While a straight line is assumed for the most energetic particles, the less energetic ones cannot have its trajectory so constrained. Therefore, the previous approximation is relaxed to allow Brownian perturbations in the transverse plane. Its propagation is described by a Green's function:
\begin{equation}
\label{eq:GFunction}
	G (x_{0+}, \mathbf{x_{0}}; L_+, \mathbf{x} | p_+ ) = \int_{\mathbf{r} (x_{0+}) = \mathbf{x_0}}^{\mathbf{r} ( L_+) = \mathbf{x}} \mathcal{D} \mathbf{r} (\xi) \exp \left\{ \frac{ ip_+}{2} \int_{x_{0+}}^{L_+} d\xi \left( \frac{ d\mathbf{r}}{ d\xiÊ} \right)^2 \right\} W (x_{0+}, L_+, \mathbf{r} (\xi) ) ,
\end{equation}
where $p_+$ is the energy of the parton that travels from $x_{0+}$ at $\mathbf{x_0}$ to $L_+$ located at the transverse coordinate $\mathbf{x}$, at the same time that its colour field is rotated by the action of the Wilson line.
\par
\begin{figure}[htbp]
	\centering
	\includegraphics[width=0.45\textwidth]{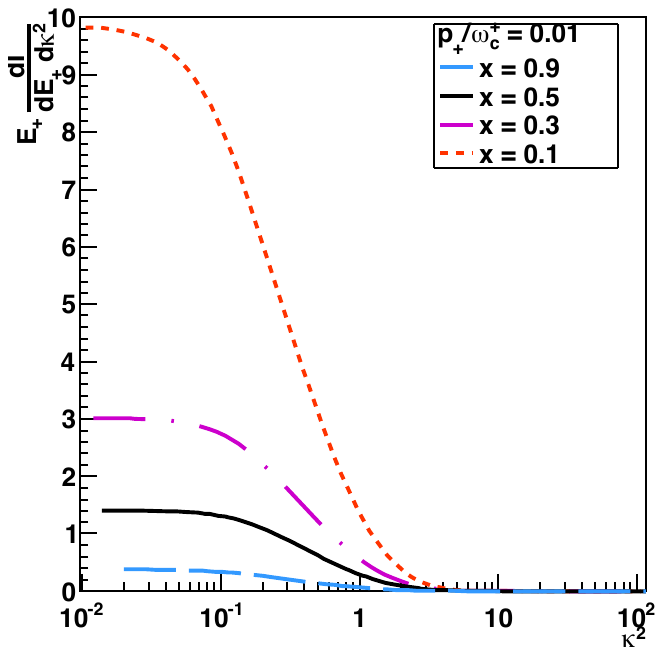}
	\includegraphics[width=0.45\textwidth]{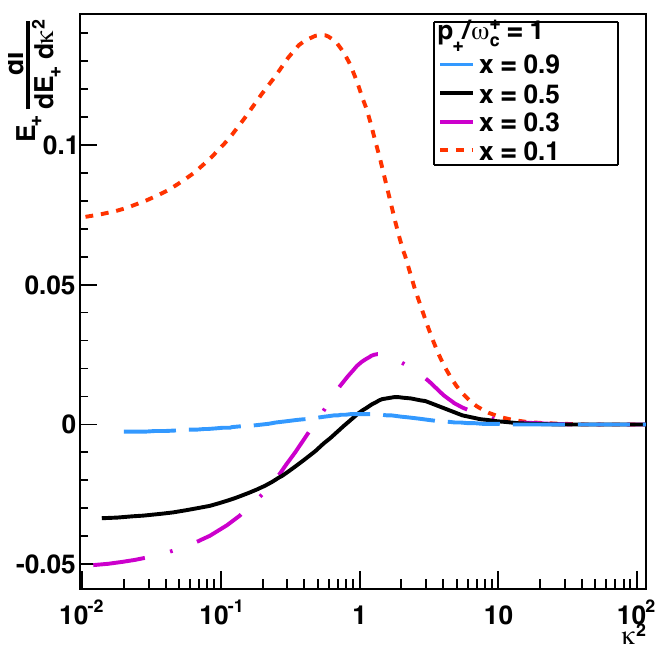}
	\caption{Evolution of the double differential medium-induced gluon radiation spectrum of a quark in a static medium for different values of $x$ and $p_+ / \omega_{c+}$.} 
	\label{fig:SpecMeu} 
\end{figure}

 Considering the $q\rightarrow qg$ splitting process, the first jet quenching calculations were performed identifying a Wilson line for the quarks and a Green's function for the radiated gluon. Recent works have tried to improve these results by going beyond soft gluon emissions. To account for finite energy corrections, an interpolation function between the soft and hard limits was computed in \cite{meu}. The resulting spectrum is shown in figure \ref{fig:SpecMeu} as a function of the dimensional variables $p_+/\omega_{c+} = 2p_+ / (\hat{q}_A L_+^2)$, $\kappa^2 = \mathbf{q}^2 /(\hat{q}_A L_+)$ with $p_+$ the initial quark energy, $L_+$ the medium length and $\hat{q}_A$ the transport coefficient that translates the squared average transverse momentum that the particle acquires per mean free path, $\lambda$. As one can observe from figure \ref{fig:SpecMeu} (left) the features already identified in previous works are recovered, namely the suppression of the radiation for large angle emissions and the absence of a collinear divergency with respect to the vacuum. Additionally, it is possible to observe the suppression of the spectrum with increasing $x$. Furthermore, the LPM effect is also observed when the medium length or density decreases (see figure \ref{fig:SpecMeu} (right)).
\par This kind of description is still limited as some of the vertex participating particles have its movement still constrained in the transverse plane. An extension beyond the eikonal propagation was derived in \cite{yacines} assuming small formation times. The considered setup includes the gluon branching process, where all particles are described by a Green's function. It was shown that independent emissions are dominant within the considered approximation. As a consequence, the showering process can be understood as a factorization of single gluon emissions, providing, in this way, a probabilistic picture that may be very useful for Monte Carlo implementation. The final result for the cross section reads:
\begin{eqnarray}
\label{eq:yacines}
	\frac{d^2 \sigma }{ d\Omega_{k_a} d\Omega_{k_b}} = 2 g^2 x (1-x) \int_{t_0}^{t_L} dt \int_{\mathbf{p}_0, \mathbf{q}, \mathbf{p}} \mathcal{P} (\mathbf{k}_a - \mathbf{p}, t_L-t) \mathcal{P} (\mathbf{k}_b - \mathbf{q} + \mathbf{p}, t_L-t) \nonumber \\
	\times \mathcal{K} (\mathbf{p} - x \mathbf{q}, x, p_{+}) \mathcal{P} ( \mathbf{q} - \mathbf{p}_0, t-t_0) \frac{ d\sigma_{hard}}{d \Omega_{p_0}} ,
\end{eqnarray}
where $x$ is the fraction of energy carried by the gluon $a$, $\mathbf{k}_a$ is final transverse momentum, $\mathbf{k}_b$ the final transverse momentum of the gluon $b$ and $\mathbf{p}(\mathbf{q})$ the initial(final) transverse momentum of the initial gluon. Equation (\ref{eq:yacines}) admits a very clean interpretation: during its propagation time, from $t_0$ to $t$, the initial gluon experience classical broadening acquiring, at the end, a transverse momentum $\mathbf{q} - \mathbf{p}_0$. Afterwards it splits quasi-instantenously into gluons $a$ and $b$ with a probability $~\alpha_s \mathcal{K} (\mathbf{p} - x \mathbf{q}, x, p_{0+})$, where each of the final gluons suffer independent broadening until $t_L$.

\section{Momentum broadening within SCET}
\label{sec3}

Other formalism to describe the energy loss as well as the momentum broadening of the shower is based on Soft-Collinear Effective Theory (SCET) \cite{scet1,scet2,scet3,scet4,scet5}. Its advantage is that the medium and the hard parton are described within the same formalism, contrary to the path-integral formalism above. This approach is based on a scale separation between the hard parton energy $Q$, the transverse momentum of the radiated gluon $l_T$ and the soft scale characteristic of the medium $T$, such that
\begin{equation}
	Q \gg l_T \gg T \Rightarrow \lambda \equiv \frac{T}{Q} \ll 1
\end{equation}
To describe the dynamics, a very high energy parton is considered with only longitudinal momenta, $p_0 = (0, Q, 0)$, that exchange some soft modes, called Glauber gluons, $p_G \sim Q (\lambda^2, \lambda^2, \lambda)$, with the medium. As a result, after each interaction, the initial parton acquires a given amount of transverse motion, $p_c \sim Q (\lambda^2, 1, \lambda)$. By re-iterating the process, a significant transverse momentum broadening is achieved. Formally, the Lagrangian of this theory is written as a function of new degrees of freedom that are given by the Glauber modes, $p_G$, collinear modes, $p_c$ and additional soft modes, $p_s \sim (\lambda^2, \lambda^2, \lambda^2)$, that characterize the medium constituents. From the Lagrangian, it is possible to identify new Feynman rules that allow to determine the medium-induced splitting kernels. The results seem to be consistent with previous ones.

\section{Colour (de)coherence}
\label{sec4}

In vacuum, the shower development is governed by the property of angular ordering between subsequent emissions. Considering a gluon emission from a quark-antiquark antenna, such effect comes from the destructive interferences that suppress the emissions at large angles. However, in the presence of a medium, while the radiation spectrum of each leg is independent and follows a medium-induced modified spectrum, the interference effects are vacuum-like and suppress the radiation at small angles instead. As a result, the total gluon spectrum off an antenna in the soft limit is dominated by the vacuum-like radiation and reads to \cite{antenna1,antenna2,antenna3} (see also \cite{antenna4} for similar results and \cite{antenna5} for a massive antenna):
\begin{equation}
	\left. dN_q^{tot} \right|_{\omega \rightarrow 0} = \frac{ \alpha_s C_F}{ \pi} \frac{ d\omega }{\omega} \frac{ \sin \theta \, d \theta }{ 1 - \cos \theta } \left[ \Theta (\cos \theta - \cos \theta_{q \bar{q}} ) + \Delta_{med} \Theta ( \cos \theta_{q \bar{q}} - \cos \theta) \right] ,
\end{equation}
where $\omega$ the energy of the radiated gluon, $\theta$ its emission angle, $\theta_{q\bar{q}}$ the cone opening angle formed by the quark-antiquark pair and
\begin{equation}
	\Delta_{med} \approx 1 - \exp \left[ - \frac{1}{12} \hat{q} \, \theta_{q \bar{q}} L^3 \right].
\end{equation}
The first contribution follows the usual angular ordering where the radiation takes place inside the cone. The second term represents the radiation that is emitted outside of the cone, following an anti-angular ordering, whose intensity is controlled by the $\Delta_{med}$ parameter. When $\Delta_{med} = 0$, there are no medium effects, and when $\Delta_{med} =1$ $(\hat{q} L \rightarrow \infty$), the emitters become completely decorrelated as a consequence of medium interactions, hence, radiating as two independent emitters.

\section{Colour Flow}
\label{sec5}

\par
\begin{figure}[htbp]
	\centering
	\includegraphics[width=0.45\textwidth]{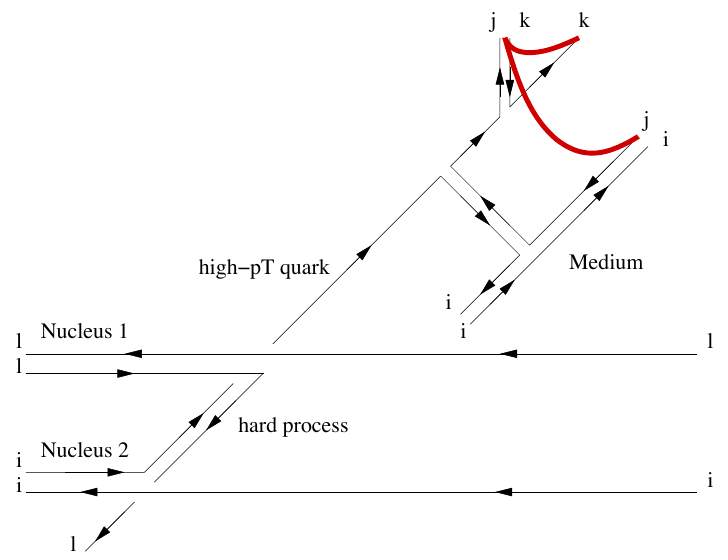}
	\includegraphics[width=0.45\textwidth]{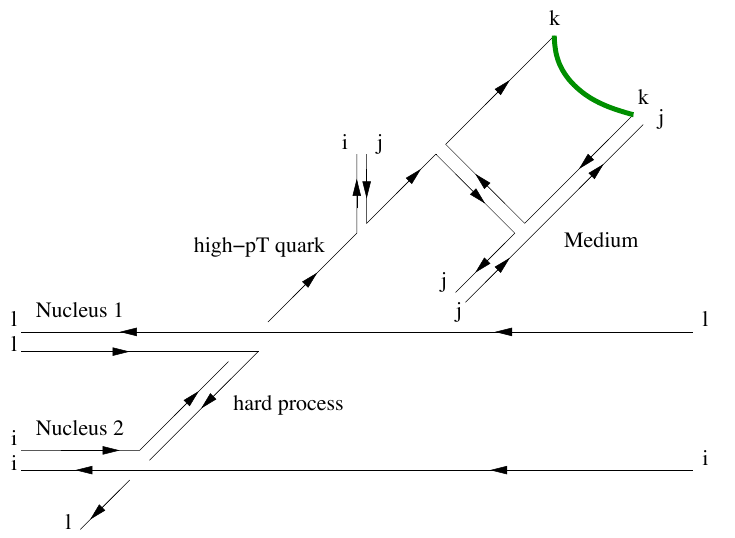}
	\caption{Example of a hard $qq\rightarrow qq$ event embedded in a nucleus-nucleus collision in which one of the high-$p_T$ quarks interacts once with the surrounding QCD matter before (left) or after (right) a gluon emission The coloured line represents the colour singlet within the Lund-string model, where the leading quark $k$ is grouped to form a cluster (figure from \cite{flow2}).} 
	\label{fig:Flow} 
\end{figure}

 After the development of the parton shower where hard partons decrease its virtuality through radiation, a hadronization prescription has to be used to cluster all coloured particles into colour-neutral hadrons. The most widely used in Monte Carlo codes are the Lund model and the singlet cluster model. In the former, implemented in PYTHIA \cite{pythia}, colour conections are made through strings that connect quarks and antiquarks, while gluons induce a kink on the string that is formed. In the latter, that is implemented in Herwig \cite{herwig}, all gluons are firstly split into quark-antiquark pairs and a clustering algorithm is applied afterwards. The works developed in \cite{flow1,flow2,flow3} address the problem of how these color flow connections are modified in the presence of a medium. Considering the setup of a gluon emission off a quark in the large $N$ limit \cite{flow1,flow2}, being $N$ the number of colours, two situations may occur: if no more interactions with the medium take place after gluon emission, this particle will remain with the leading fragment and no modification with respect to the vacuum should be observed (see figure \ref{fig:Flow} (left)). Nonetheless, if there is an interaction with the medium after the gluon is emitted, the color flow is broken by this interaction. The gluon decoheres from the string and its energy is lost from the leading fragment (figure \ref{fig:Flow} (right)), inducing an extra-amount of energy loss with respect to a standard jet quenching formulation with a fixed $\hat{q}$ parameter. The decohered partons will then hadronize into soft components.

\section{Summary}
\label{sec6}

There has been a lot of activity in these last years to improve the state-of-the-art jet quenching theory. The main objective is to build a complete picture of the shower branching process when in the presence of a medium with the characteristics of a QGP.  Among the several developments, in this manuscript I addressed kinematic-related improvements, where the gluon emission spectrum was calculated to account for finite energy loss, instead of only soft radiation, and the eikonal approximation was relaxed to allow Brownian perturbations in the transverse plane of the propagating particles. This was computed within two different formalisms: the path-integral one and the SCET. With respect to the parton shower evolution,  a probabilistic picture for the shower development was identified, valid for the approximation of infinite medium length. The works concerning the colour correlation between successive emitters within an antenna setup were also briefly mentioned, where it is possible to identify the relevant scales that allow for a geometrical separation of the radiation. In the presence of a medium, large angle radiation is not suppressed as in vacuum leading to large angle soft emissions. Finally, the main results from the studies of the modification of colour connections inside a hot and dense medium show that an extra amount of energy loss can be induced through the breaking of the leading string by medium interactions. All these achievements are relevant for particle and jet observables at nucleus-nucleus collisions and the phenomenological use of these new theoretical ingredients is currently under development.

\acknowledgments

{\small I thank N. Armesto, J. G. Milhano and C. A. Salgado for the useful discussions and for carefully reading this manuscript. I acknowledge the financial support from the portuguese Funda\c{c}\~{a}o para a Ci\^{e}ncia e Tecnologia grant SFRH /BD/64543/2009 and project CERN/FP/123596/2011,  and by the European Research Council grant HotLHC ERC-2011-StG-279579.}

\end{document}